# Large room temperature anomalous Nernst effect coupled with topological Nernst effect from incommensurate spin structure in a Kagome antiferromagnet

Jiajun Ma[1], Jiaxing Liao[1], Yazhou Li[1], Yuwei Zhang[1], Jialu Wang[1], Jinke Bao[1], Yan Sun[2,#], Shuang Jia[3,*], Yuke Li[1,*]

1. School of Physics and Hangzhou Key Laboratory of Quantum Matters, Hangzhou Normal University, Hangzhou 311121, China

2. Shenyang National Laboratory for Materials Science, Institute of Metal Research, Chinese Academy of Sciences, Shenyang, 110016, China

3. International Center for Quantum Materials, School of Physics, Peking University, Beijing 100871, China

## Abstract

**Kagome magnets exhibit a range of novel and nontrivial topological properties due to the strong interplay between topology and magnetism, which also extends to their thermoelectric applications. Recent advances in the study of magnetic topological materials have highlighted their intriguing anomalous Hall and thermoelectric effects, arising primarily from large intrinsic Berry curvature. Here, we report observation of a large room-temperature (RT) anomalous Nernst effects (ANE) of $S_{xy}^A \sim 1.3~\mu V~K^{-1}$ in the kagome antiferromagnet (AFM) ErMn$_6$Sn$_6$, which is comparable to the largest signals observed in known magnetic materials. Surprisingly, we further found that a significant topological Nernst signal at RT and peaking a maximum of approximately 0.2 $\mu V~K^{-1}$ at 180 K, exactly coupling with ANE in the spiral AFM state, originates from the real-space nonzero spin chirality caused by incommensurate spin structure. This study demonstrates a potential room-temperature thermoelectric application platform based on Nernst effect, and provides insights for discovering significant anomalous and topological transverse transport effects in the incommensurate AFM system.**

## Introduction

Kagome lattice materials exhibit exotic electronic structures, such as flat bands, Dirac/Weyl points, and van Hove singularities, making them an ideal platform for investigating anomalous quantum transport phenomena[1-10]. These unique properties hold significant promise for advanced applications, such as ultrafast electronics, spintronics, high-density data storage, and thermoelectric devices[11-14]. In recent decades, conventional thermoelectric energy harvesting based on the Seebeck effect has advanced rapidly. However, its limitation requires precisely engineered p-n junction architectures to generate considerable power output[15-17]. Conversely, transverse thermoelectric effects, particularly anomalous Nernst effect (ANE), can significantly reduce device fabrication (Figure 1a) by eliminating additional resistivities from multiple electrical connections and interfaces[9, 18]. This allows monolithic thermopile arrays to be realized within a single material (Figure 1a). However, practical applications demand materials that simultaneously exhibit room temperature magnetic order and large anomalous transverse thermoelectric coefficients, a combination rarely satisfied in existing systems.

Recent years, the Mn-based Kagome magnet $RMn_6Sn_6$ (R=Gd-Lu, Y, Sc) family[3, 7, 19] has been found to exhibit Chern gaps in their electronic structure. As the radius of R ion decreases, the gap size gradually narrows. This trend is accompanied by an increasing competition between R-Mn and Mn-Mn interlayer exchange interactions, making this system an excellent platform for exploring anomalous transport phenomena. The compounds $RMn_6Sn_6$ (R= Gd-Ho) stabilize an out-of-plane ferrimagnetic (FIM) order driven by strong R–Mn coupling, giving rise to a sizable anomalous Hall effect (AHE). In contrast, the ones with non-magnetic R = Tm, Lu, Y and Sc favor AFM or helimagnetic coupling between Mn layers, leading to the emergence of a topological Hall effect (THE). Especially, $ErMn_6Sn_6$ (EMS) exactly locates at the boundary between the incommensurate AFM and the FIM phase, making it simultaneously possessing the FIM ground state and the incommensurate AFM phase along c-axis in different temperature regimes[3, 20-23], as illustrated in Figure 1b. Consequently, it exhibits an anisotropic THE with opposite signs associated with anisotropic magnetic structures[24], and undergoes a transition from THE to AHE with magnetic field along ab-plane[25]. Based on these advances, EMS, a system characterized by its incommensurate spin textures, could be considered as a unique system to explore the exotic quantum transport properties within those magnetic regimes.

Moreover, a large ANE $S_{xy}^A$ (~ $Q_sM_s$), scaling with magnetization, is usually realized in ferromagnetic materials due to a large Berry curvature near the Fermi level. Intriguingly, significant ANE has also been reported in several AFM systems with noncollinear spin textures, such as $Mn_5Si_3$[26], MnGe/Si[27, 28], $Mn_3X$(X=Sn, Ge)[29-31], FeGe[32], $YbMnBi_2$[33], and $MnBi_2Te_4$[34]. Despite these promising candidates, most of known AFM materials show relatively small ANE values or require low temperatures, severely limiting their practical thermoelectric applications. Recent breakthroughs in AFM $RMn_6Sn_6$ (R= Er, Y and Sc) and FIM $TbMn_6Sn_6$ have demonstrated large room-temperature ANE[35-38], revitalizing interest in kagome lattices as a platform for high-performance transverse thermoelectric devices. Furthermore, the intrinsic geometric frustration of the kagome spin lattice often stabilizes nontrivial spin textures, which may give rise to an additional quantum phenomenon—topological Nernst effect (TNE). The TNE is notoriously difficult to isolate experimentally as it often overlaps with anomalous Nernst signals. So far, TNE is thus less reported and its microscopic origin in AFM kagome systems remains poorly understood. For example, the interplay between Chern-gap and unique magnetic orders in $RMn_6Sn_6$ [3], nonzero spin chirality of the skyrmion bubble phase in $Fe_3Sn_2$ [39], as well as noncollinear AFM ordering in $Mn_3Sn/Ge$[29-31], suggests a rich yet unresolved mechanism underlying TNE.

Here, we report the coexistence of a large room temperature ANE and TNE in the distorted triplet spiral (DTS) AFM state along the z-axis in EMS. Large $S_{xy}^A$ reaches a substantial value of 1.38

$\mu V~K^{-1}$ at 330 K, comparable to those observed in the known topological AFMs. The prominent ANE is attributed to strong Berry curvature in momentum space, arising from Chern gap Dirac fermions[29, 38, 40-42]. Surprisingly, a significant TNE is found at RT, peaking at approximately 0.2 $\mu V~K^{-1}$ around 180K, which is comparable to other known topological magnets. This TNE, which couples with ANE, is ascribed to a non-zero scalar spin chirality resulting from incommensurate spin textures related to the z-axis DTS AFM order.

## I. EXPERIMENTAL DETAILS

EMS single crystals were grown using a self-flux method. High-quality elemental Er, Mn, and Sn powders were mixed in a precise ratio of 1:6:30 and placed within a high-vacuum quartz tube. The quartz tube was heated to 1000°C and held this temperature for 24 hours in a box furnace. Subsequently, the furnace was cooled down to 950°C at a rate of 6°C/h. We then heated the tube to 990 °C and subsequently cooled it down to 600 °C at a cooling rate of 6 °C per hour. Finally, the single crystals can be obtained using a centrifuge to remove excess Sn. The size for the grown hexagonal EMS plates is about 3mm×2mm×0.4mm.

The parameters of EMS crystal structure are identified by the X-ray diffraction with CuK$_\alpha$ radiation at room temperature, and its stoichiometry was measured by an energy-dispersive X-ray spectroscopy, as seen in Figure S1. Magnetization measurements were conducted using a Quantum Design MPMS7 instrument. Magneto-electrical and thermoelectric transport measurements were performed in a Cryogenic-14 T magnet system. The resistivity and Hall resistivity were measured using a standard six-probe technique, and the thermopower and Nernst effect measurements were performed using a self-designed sample puck with one-heater (2.2 kΩ resistor)-two-type-E (Chromel-Constantan) thermocouples. The configuration of those measuring setup is shown in Figure 1a. The temperature difference $\Delta T$ between the two thermal meters was less than 2.4 % of the average sample temperature. The voltage/current signals were collected using the Keithley-2182A nanovoltmeters and the Keithley-6221 current source, respectively.

## II. RESULTS AND DISCUSSION

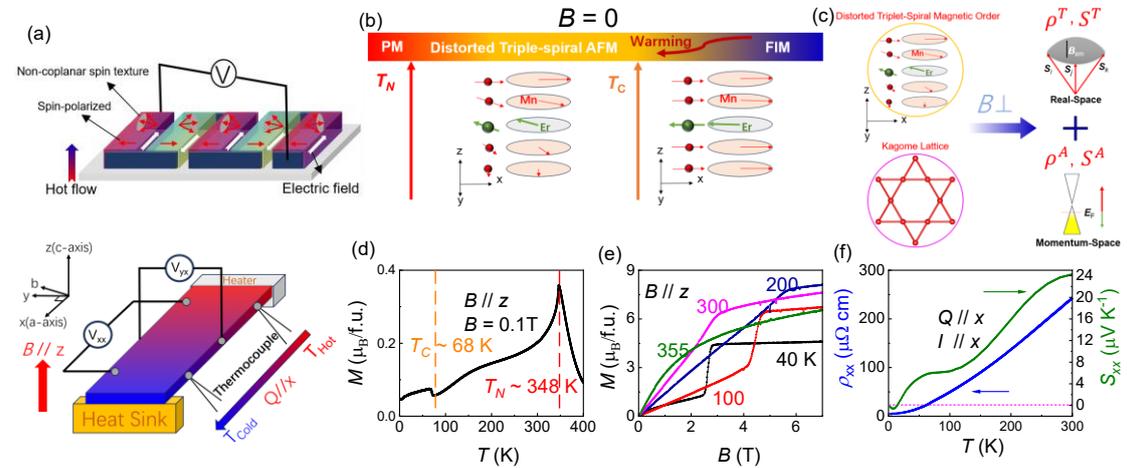

Figure 1 (a) A schematic picture of a thermoelectric module and the configuration of longitudinal and transverse thermoelectric transport measurements. (b) The schematic diagram of magnetic structure evolution with temperature for EMS. (c) A cartoon illustrating the relationship between spiral magnetic orders and anomalous/topological transport responses in EMS. (d) Temperature dependence of magnetization as $B \parallel z$-axis. (e) The isothermal magnetization curves from 40 to 355 K. (f) Resistivity (left) and thermopower (right) as functions of temperature for heat/current along the x-axis.

The magnetization curve $M(T)$ with zero-field-cooled mode at $B = 0.1$ T along the z-axis in Figure 1d demonstrates that EMS undergoes two successive magnetic phase transitions at $T_N = 348$ K and $T_C = 68$ K (Figure S2), respectively. Neutron scattering experiments[21] have confirmed that the former corresponds to the transition from paramagnetic state into a spiral order, while the latter marks the appearance of FIM state[22, 24] due to its competing Er-Mn and Mn-Mn couplings along the $c$-axis. In the spiral AFM state, a field-induced phase transition from an AFM to FIM state occurs within the range of 2.2-5.5 T (Figure 1e), consistent with the presence of in-plane magnetic moments[25], as depicted in Figure 1b. The resistivity $\rho_{xx}$ decreases nearly linear with temperature cooling down, dropping from 243.4 $\mu\Omega$.cm at 300 K to 4.8 $\mu\Omega$.cm at 4 K (Figure 1f). This yields a large Residual Resistivity Ratio (RRR = $\rho_{300K}/\rho_{4K}$) of 51, indicative of the high quality of single crystals. The thermopower $S_{xx}$ is positive and decreases monotonically with temperature in Figure 1f, suggesting a majority of hole-type carriers. Below 15 K, $S_{xx}$ crosses zero (marked by pink dashed line) and changes sign from positive to negative. While this signals a crossover in the dominant carrier type from holes to electrons, yet it does not necessarily imply a shift in the Fermi level. The large $S_{xx}$ reaches 24 $\mu$V K$^{-1}$ at 300 K, a value comparable to that of other Kagome materials such as TbMn$_6$Sn$_6$[38], YMn$_6$Sn$_6$[36], and ScMn$_6$Sn$_6$[37].

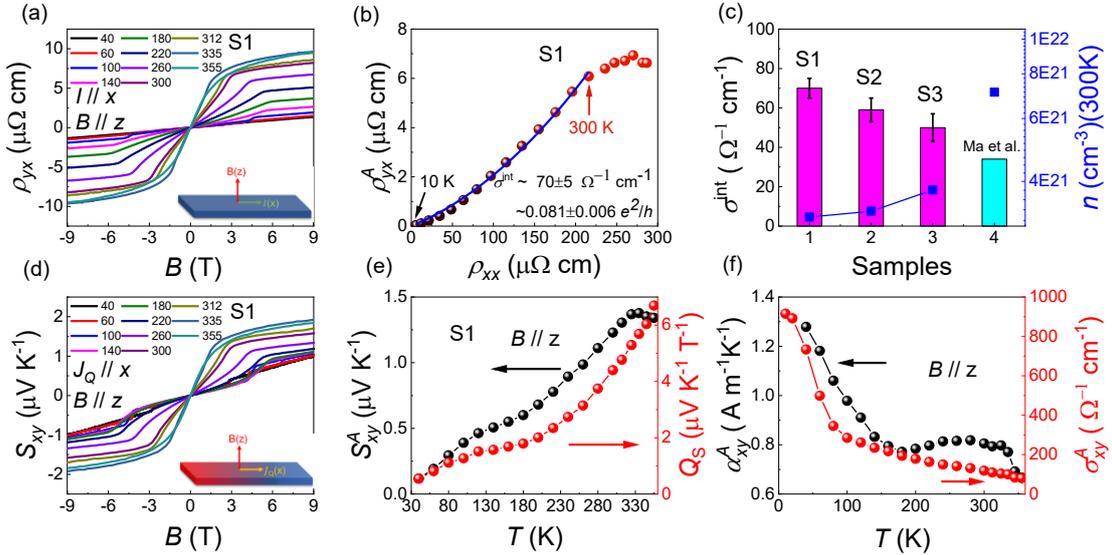

Figure 2 For $B \parallel z$-axis (a) Hall resistivity $\rho_{yx}$ and (d) Nernst signal $S_{xy}$ as functions of magnetic fields from 40 to 335 K. (b) The anomalous Hall resistivity $\rho_{yx}^A$ against $\rho_{xx}$ from 10 K to 355 K. The bule line represents the polynomial fitting of the data, which gives the intrinsic anomalous Hall conductivity $\sigma_{xy}^{int} = (0.081 \pm 0.006)$ e$^2$/h per manganese kagome layer. (c) The intrinsic Hall conductivity vs. different samples. (e) The ANE $S_{xy}^A$ (left), and the anomalous Nernst coefficient $Q_s$ (right) as functions of temperature. (f) Temperature dependence of the anomalous transverse thermoelectric conductivity, $\alpha_{xy}^A$, and anomalous Hall conductivity, $\sigma_{xy}^A$.

Figure 2a and 2d present Hall resistivity $\rho_{yx}$ and Nernst signal $S_{xy}$ of S1 as functions of the magnetic fields along the z-axis from 40 K to 355 K. Both $\rho_{yx}$ and $S_{xy}$ show a sudden jump at a critical magnetic field, followed by a linear increase at higher fields, closely resembling the isothermal magnetization curves $M(B)$ for $B$ // z-axis (See FigureS2). The correspondence suggests that anomalous Hall and Nernst components dominate, which are likely proportional to the magnetization. Similar reports have been found in conventional FMs such as Fe[43], $Fe_3O_4$[44], as well as in typical topological magnets like $Co_3Sn_2S_2$[9, 45], and $Co_2MnGa$[46]. Furthermore, by subtracting the linear terms (Normal Hall/Nernst effect), the anomalous Hall and Nernst components $\rho_{yx}^A$ and $S_{xy}^A$ are extracted in Figure 2b and 2e. To determine the intrinsic Berry curvature contribution, the $\rho_{yx}^A(T)$ from 10 to 300 K can be fitted using the following equation[47]: $\rho_{yx}^A = \sigma_{xy}^{int}\rho_{xx}^2 + \beta^{skew}\rho_{xx}$, where $\sigma^{int}$ is the intrinsic anomalous Hall conductivity (AHC), and $\beta^{skew}$ denotes the skew scattering contribution parameter. The obtained intrinsic AHC (Figure 2c) $\sigma_{xy}^{int}$ is ~ 70 $\Omega^{-1}$ $cm^{-1}$ for S1, and 50-60 $\Omega^{-1}$ $cm^{-1}$ for S2 and S3 (see Figure S3 and S4 in SI). This corresponds to a contribution of 0.081 $e^2/h$ per kagome layer for S1, which is approximately twice that reported in the previous studies[3]. This enhancement can be attributed to the closer proximity of the Fermi level to the Chern gap in our sample, which induces a larger Berry curvature in momentum space. Additionally, the carrier concentration (Figure 3c) is almost an order of magnitude lower than values reported in previous works[3], further supporting this interpretation (Figure S3). The AHC $\sigma_{xy}^A = \frac{\rho_{yx}^A}{\rho_{xx}\rho_{yy}+(\rho_{yx}^A)^2}$, as plotted in Figure 2f, exhibits a monotonic increase with a significant rise below 100 K, reaching ~ 900 $\Omega^{-1}$ $cm^{-1}$ at 10 K. This value is remarkably close to that of the intrinsic FM Weyl semimetal $Co_3Sn_2S_2$ (~1130 $\Omega^{-1}$ $cm^{-1}$)[5, 9].

The magnitude of $S_{xy}^A$ increases monotonically with temperature in Figure 2e, peaking a value of 1.38 $\mu V$ $K^{-1}$ at 335K. This value is comparable to those reported for other topological ferromagnets such as $Co_3Sn_2S_2$[9], $Fe_3Sn_2$[42] and $Co_2MnGa$[46]. Notably, the Nernst signal persists above the $T_N$, suggesting the existence of spin fluctuations in EMS[48], consistent with the magnetization data (see Figure S2). Using the relation $S_{xy}^A = Q_s\mu_0 M$, the anomalous Nernst coefficient $Q_s$ is calculated and plotted on the right axis of Figure 2e. At 355 K, $Q_s$ reaches a value of 6.5, significantly exceeding those of conventional ferromagnets, for which $Q_s$ typically falls within between 0.05 and 1. This enhanced $Q_s$ highlights the substantial contribution of Berry curvature to the ANE in the EMS system. The anomalous thermoelectric conductivity $\alpha_{xy}^A$ (Figure 2f) can be given by the expression: $\alpha_{xy}^A = \frac{\rho_{xx}S_{yx}^A - \rho_{yx}^A S_{xx}}{\rho_{xx}\rho_{yy}+(\rho_{yx}^A)^2}$, where $\rho_{xx}$ and $S_{xx}$ denote the longitudinal resistivity and thermopower, respectively, while $\rho_{yx}^A$ and $S_{yx}^A$ represent the anomalous Hall and Nernst signal. $\alpha_{xy}^A$ attains a value of 1.3 A $m^{-1}$ $K^{-1}$ at 40 K, which is comparable to those reported in prominent topological magnets, including $Co_2MnGa$[46], $Fe_3X$-family[49], and $Fe_3Sn_2$[39, 42].

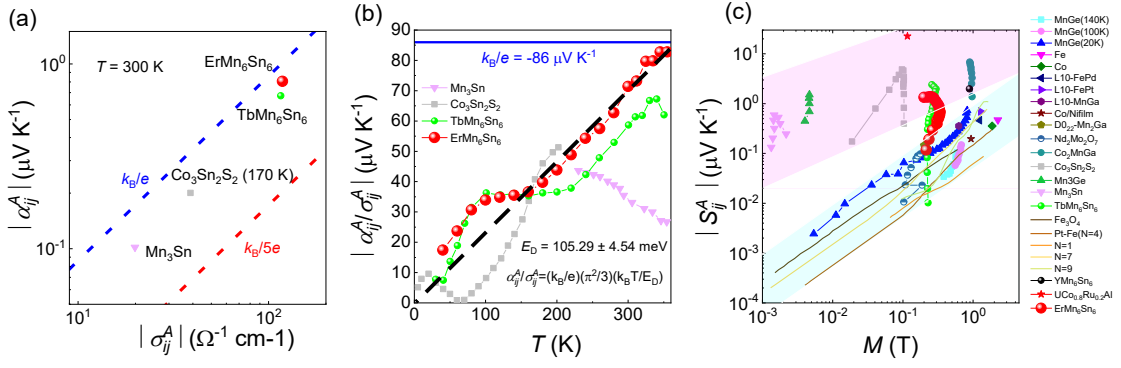

Figure 3 (a) Room-temperature $\alpha_{ij}^A/\sigma_{ij}^A$ for various magnetic materials including EMS falls within the range between $k_B/e$ (red line) and $k_B/5e$ (blue line). (b) The $\alpha_{ij}^A/\sigma_{ij}^A$ ratio as a function temperature for typical kagome magnets. Here, EMS approaches 86 µV K$^{-1}$ at room temperature. (c) Magnetization dependence of the ANE for other ferromagnets and EMS.

Within the intrinsic framework, as shown in Figure 3a-3b, this ratio $|\alpha_{ij}^A/\sigma_{ij}^A|$ in EMS approaches the theoretical value of 86 µV K$^{-1}$ and lies between $k_B/e$ and $k_B/5e$ near room temperature when the thermal de Broglie wavelength in the plane perpendicular to the magnetic field becomes comparable in magnitude to the Fermi wavelength under ambient conditions[41]. Such behavior aligns with observations in most discovered topological magnets, such as Co$_3$Sn$_2$S$_2$[40], and TbMn$_6$Sn$_6$[38], and confirms the intrinsic anomalous Hall/Nernst effect induced by a nontrivial Berry curvature. Based on the massive Dirac model's Mott formula[38], the fitted energy scale $E_d \sim 105$ m eV in EMS falls into between Fermi energy and Chern gap (Figure 3b). Below approximately 160 K, $\alpha_{ij}^A/\sigma_{ij}^A$ deviates from the linear temperature-dependent, similar to that observed in TbMn$_6$Sn$_6$[38], likely indicating the enhancement of spin excitations due to the intensified competition between the magnetic crystalline anisotropy energies of Er and Mn. A comparison of the ANE $S_{yx}^A$ in EMS with conventional FMs and topological magnets is shown in Figure 3c. $S_{yx}^A$ in EMS is nearly an order of magnitude larger than those observed in conventional FMs (within the blue shaded region), and is comparable to the values reported in TbMn$_6$Sn$_6$[38], Co$_3$Sn$_2$S$_2$[9], and Co$_2$MnGa[41]. These findings highlight the significant role of large Berry curvature, induced by the interplay of Chern gap and spiral AFM orders, in significantly enhancing the ANE[38].

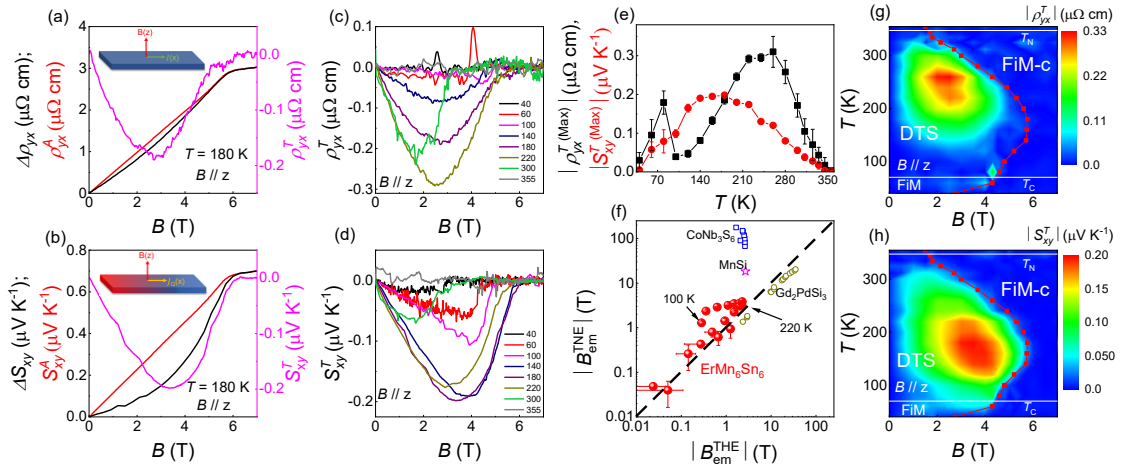

Figure 4 (a), (b) The topological Hall $\rho^T$ and topological Nernst effect (pink lines) $S^T$ are extracted by subtracting the anomalous term (black lines) that is proportional to the magnetization from the sum of topological and anomalous Hall/Nernst components (red lines), $\Delta\rho_{xy}, \Delta S_{xy}$, at 180K as B // z-axis. (c) and (d) represent the field-dependent THE and TNE from 40 to 355 K. (e) Temperature-dependent maximum values of THE and TNE. (f) The emergent effective magnetic fields derived from THE (black) vs. TNE (red). Phase diagram of EMS and contour plot of THE (g) and TNE (h) as functions of magnetic fields and temperature. The red curve represents the critical field extracted from the magnetization curve (M vs B).

In addition to the large ANE, another intriguing phenomenon observed in EMS is the TNE. The total Nernst signal can be expressed as: $S_{NE} = S_{xy}^o + S_{xy}^A + S_{xy}^T$, where $S_{xy}^o = S_0 H$ represents the ordinary Nernst signal, $S_{xy}^A = Q_s \mu_0 M$ corresponds to the ANE, and $S_{xy}^T$ refers to the topological Nernst signal. The ordinary and anomalous Nernst signal can be combined as $S_{NE} = S_0 H + Q_s 4\pi M$. The slope $S_0$ and intercept $Q_s 4\pi$ can be estimated from the linear plot of $S_{ij}$/M vs. $B$/M in the high-field region. Consequently, the topological Nernst signal $S_{xy}^T$ can be extracted by subtracting $S_{xy}^A + S_{xy}^o$ from $S_{NE}$ (Figure S7). The Figure 4b illustrates the three different components of Nernst signal at 180 K: the difference $\Delta S_{xy} = S_{NE} - S_{xy}^o$, anomalous Nernst term $S_{xy}^A$, and topological Nernst term $S_{xy}^T$. The $S_{xy}^T$ reaches a maximum value of ~ 0.2 $\mu$V K$^{-1}$ around 4 T and 180 K. Above $T_N$ or below $T_c$, the $S_{xy}^T$ sharply declines and eventually vanishes in Figure 4d and 4e. The topological Hall resistivity (THR) $\rho_{yx}^T$ shows behavior similar to that of $S_{xy}^T$ in Figure 4a-4e. The THR attains a maximum value of -0.31 $\mu\Omega$ cm at 260 K(Figure 4b), much larger than the previously reported values (e.g. -0.2$\mu\Omega$ cm at $T$ = 200 K)[24]. As temperature deviates from 260 K, either by cooling or warming, the THR begins to decrease. Notably, the $\rho_{yx}^T$ changes sign from negative to positive below 100 K, with positive THR emerging only within a narrow magnetic field range (3.5-4.5 T) and vanishing below 40 K. A possible explanation involves the magnetic phase transition from the DTS to an in-plane FIM order ~ 70 K. In the resulting FIM state, an applied magnetic field can induce spin reorientation, giving rise to non-trivial spin textures that reverse the chirality of the spin structure and consequently lead to the sign reversal of the THE. A similar phenomenon has been reported in the skyrmion magnet MnGe[50].

The temperature-dependent peak values of $S_{xy}^T$ and $\rho_{yx}^T$ are summarized in Figure 4e, showing that their maxima occur at distinct temperatures. This difference arises from their distinct physical origins: the THE (reflected in $\rho_{yx}^T$) measures the transverse deflection of charge carriers and is governed by the Berry curvature integrated over all occupied bands, whereas the TNE (reflected in $S_{xy}^T$) probes the transverse flow of entropy and is sensitive to the Berry curvature near the Fermi energy. The lower peak temperature of TNE (180 K) indicates that above this temperature, thermal broadening suppresses the entropy-driven response of $S_{xy}^T$ even though the underlying spin textures remain present. Prior studies have similarly shown that the AHE and ANE exhibit distinct temperature-dependent peak positions, a behavior consistently observed in other topological magnets, such as Co$_3$Sn$_2$S$_2$[9, 40], CoMnGa$_2$[51], and Fe$_{3-x}$GeTe$_2$[52]. Their magnitudes are significantly larger than those observed in several typical topological magnets, including canted AFM FeGe[8] and noncollinear AFM NdMn$_2$Ge$_2$[53], Fe$_3$Sn$_2$[39] and the skyrmion lattice MnSi[28], all of which exhibit topological Hall/Nernst effect.

The Berry curvature in momentum space, which acts as an effective emergent magnetic field, and breaks time-reversal symmetry, is responsible for the intrinsic AHE and ANE. In a parallel manner, topological spin textures such as those arising from skyrmion lattices[28], scalar spin chirality[54], can also produce non-zero Berry curvature. This curvature manifests a real space emergent magnetic field $B_{em}(r)$ for conducting electrons, giving birth to the THE[54, 55 ,57] and TNE[56, 58]. Here, we derive the emergent effective magnetic field $B_{em}(r)$ for both effects using the respective formulas $\rho_{yx}^T = R_0 B_{em}^{THE}(r)$, $S_{xy}^T = \nu B_{em}^{TNE}(r)$, as shown in Figure 4f. The consistent difference of nearly 2 T between $B_{em}^{THE}$ and $B_{em}^{TNE}$ from 100 K to 220 K suggests that the influence of the band structure on THE and TNE cannot be completely ruled out, especially when the sizes of magnetic textures approach the lattice spacing. A similar phenomenon has been observed in recent studies of the metastable skyrmion phase in MnSi[28] and the chiral antiferromagnet $CoNb_3S_6$[58]. Further research on the true reason of these effects should be performed in the future.

The phase diagram of THR and TNS as functions of magnetic field and temperature is presented in the Figure 4g and 4h, respectively. As the out-of-plane magnetic fields rise, the magnetic structure of EMS undergoes a transition from a triple-spiral to a collinear ferrimagnetic state. Notably, both types of topological responses are predominantly observed within the spiral magnetic order, suggesting that the spins arrange in a spiral pattern with a non-uniform rotation angle along the c-axis, breaking the time-reversal symmetry and generating a real-space emergent magnetic field for conducting electrons. This emergent magnetic field, analogous to the Berry curvature in momentum space, contributes to the topological transverse transport effects. In addition, a distinctive feature of $ErMn_6Sn_6$ is the coexistence of the THE and TNE at low magnetic fields. This behavior stands in sharp contrast to $YMn_6Sn_6$[54, 59], which only exhibits the THE under high fields, and to $TbMn_6Sn_6$[38], which does not display the THE. Furthermore, the TNE and THE are strongly coupled with the AHE/ANE within its incommensurate AFM phase, suggesting a unified origin of these transverse responses in the spiral magnetic state. Additionally, although our results do not directly clarify the precise mechanism by which out-of-plane magnetic field-driven chiral triple-spiral magnetic order generates non-collinear magnetic structures, we have quantitatively mapped the distributions of the THE and TNE across a wide range of temperatures and magnetic fields. This comprehensive mapping serves as a valuable benchmark for future investigations into the unique spin textures of $ErMn_6Sn_6$ using advanced techniques such as neutron scattering and muon spin spectroscopy.

## III. CONCLUSION

In summary, we systematically investigated the transverse thermoelectric properties of the kagome antiferromagnet $ErMn_6Sn_6$ and reported the coexistence of a large room-temperature ANE and TNE in its distorted triplet spiral AFM state. This anomalous Nernst signal reaches the maximum of ~ 1.38 $\mu V\ K^{-1}$ at 330 K, comparable to values reported in a few well-known topological magnets. This large ANE originates from strong Berry curvature in the momentum space, induced by the Chern gap Dirac fermions. A significant topological Nernst signal at RT and peaking at approximately 0.2 $\mu V\ K^{-1}$ around 180 K, is observed, which couples with the ANE. This TNE is ascribed to the non-zero scalar spin chirality generated by incommensurate DTS AFM spin texture along the z-axis, which generates a real-space emergent magnetic field for conducting electrons.

Furthermore, ANE and TNE are strongly coupled across a broad temperature window from 40 to 335 K, which is corroborated by the strong correlation between the temperature dependence of the ANE/TNE signals and the scalar spin chirality/Berry curvature. Our findings demonstrate that incommensurate spin textures in the Kagome system simultaneously host substantial TNE and ANE at RM, providing a promising route for exploring novel transverse thermoelectric phenomena. Future research on the optimization of spin structures, crystal qualities, and device integration will further enhance the performance of kagome magnets, paving the way for their widespread application in energy harvesting and thermal management.

Corresponding authors: sunyan@imr.ac.cn; gwljiashuang@pku.edu.cn; yklee@hznu.edu.cn

## Acknowledgements

This research was supported in part by the National Natural Science Foundation of China (under Grants No. U1932155, 52271016, 12225401, 12404047), Y. Li acknowledges the Hangzhou Joint Fund of the Zhejiang Provincial Natural Science Foundation of China (under Grants No. LHZSZ24A040001), and the HZNU scientific research and innovation team project (No. TD2025013). J.-K. B. acknowledges support from Beijing National Laboratory for Condensed Matter Physics (Grant No. 2023BNLCMPKF019)

## Conflict of Interest

The authors declare no conflict of interest.

## Keywords

Antiferromagnetism, Kagome magnets, Nernst effect, Incommensurate spin textures